# Meissner levitation of a millimeter-size neodymium magnet within a superconducting radio frequency cavity

N. K. Raut, J. Miller, J. Pate, R. Chiao, and J. E. Sharping

*Abstract*— We report on the magnetic levitation of a millimeter-sized neodymium permanent magnet within the interior of a superconducting radio frequency (SRF) cavity. To the best of our knowledge, this is the first experimental work on levitating a magnet within an SRF cavity. The cavity is a coaxial quarter-wave microwave resonator made from 6061 aluminum, having a resonance frequency of 10 GHz and a loaded Q of 1400. The cylindrical magnet (N50) has a height of 1 mm, a diameter of 0.75 mm, a mass of 4 mg, and a remanence of 1.44 T. This produces a peak magnetic field 140 times greater than the critical field of aluminum. The magnet is placed either on the top of the coaxial portion of the cavity or on the cavity floor before cooling it below the superconducting transition temperature of aluminum. The coaxial mode's resonance frequency shifts as a function of the levitation height of the magnet and gives an idea of the magnet's position and mechanical motion. We observe a transition at a temperature of 650 mK where the Meissner effect levitates the magnet as the material beneath the magnet becomes superconducting. The magnet is levitated to a height of 2.5 mm above the surface of the cavity stub, which is a sufficient separation for the field strength of the magnet at the surface of the stub to be less than the critical field strength of the superconducting aluminum. We measure a 120 MHz upshift in the cavity resonance as the magnet is levitated from the top of the stub and 15 MHz downshift as it levitates from the floor of the cavity. Our measurements are consistent over several heating and cooling cycles. Our work provides a path towards a novel optomechanical system.

## I. Introduction

Superconducting radio frequency (SRF) cavities are promising for use in cavity optomechanics and quantum information processing due to their well-defined frequency spectrum, low loss and electric and magnetic field localization properties [1-4]. Besides SRF cavities, various architectures have been developed to strongly couple an object's mechanical modes with electromagnetic modes [5-11]. In the above examples, the mechanical object is physically tethered in some way to a boundary, and one can consider levitating the object so that the tether can be removed. Electromagnetic to mechanical coupling has been observed in optically levitated systems [12-20]. Here, the intense optical field is used to trap the dielectric particle so that there is less dissipation arising from clamping. and reduced thermal contact [21]. The main challenges become maintaining stable trapping at high vacuum and reducing the mechanical trapping noise arising from photon recoil and heating [22-23]. Theoretical and experimental studies of magnetic levitation suggest that it may be a good alternative to optical trapping [24-27] in electro-mechanical systems. The work on stable magnetic levitation over a type II superconductors is well established [28-30], and the basis for a common classroom demonstration. However, levitation above a type II superconductor exhibits pinning of flux vortices arising from the magnetic field at the surface of the superconductor. Flux pinning introduces dissipation in the system, but the Meissner state in a type I superconductor does not exhibit flux vortices, leading to more efficient mechanics [31]. In recent years, a stable levitation of a permanent magnet above a punctured type I superconductor was demonstrated for use in optomechanical experiments [32], [33].

In this paper, we report field-cooled magnetic levitation of a millimeter-scale neodymium magnet within a cm-scale aluminum coaxial-stub SRF cavity. The SRF cavity has a resonance of 10.4 GHz and a loaded Q of 1400, and is fabricated from 6061 (97.9% pure) aluminum. The cylindrical (0.75-mm diameter by 1-mm high) neodymium magnet has a maximum magnetic field of 1.4 T, which is ~140-times greater than the critical field of the aluminum. Below the critical temperature for superconductivity, the magnet levitates 2.5 mm above the surface of the material and its behavior is monitored by measuring changes in the SRF resonant frequency. Room temperature measurements and simulations of magnet position's effect on the cavity mode support our conclusions. This novel magnet-cavity system provides a means to couple the low-frequency

---

(Style: TAS First page footnote) Manuscript receipt and acceptance dates will be inserted here. Acknowledgment of support is placed in this paragraph as well. Consult the IEEE *Editorial Style Manual* for examples. This work was supported by the IEEE Council on Superconductivity under contract. ABCD-123456789. *(Corresponding author: Jay E. Sharping.)*

N. K. Raut is with University of California, Merced, CA 95343 (e-mail: nraut@ucmerced.edu).
J. Miller is with University of California, Merced, CA 95343 (e-mail: jmiller@ucmerced.edu).
J. Pate was with University of California, Merced, CA 95343 (e-mail: jpate@ucmerced.edu).
R. Chiao was with University of California, Merced, CA 95343 (e-mail: rchiao@ucmerced.edu).
J. E. Sharping is with University of California, Merced, CA 95343 (e-mail: jsharping@ucmerced.edu).

mechanical motion of the magnet with other objects whose quantum states can be probed and manipulated, such as magnons and transmons [3]. Consequently, the system is a promising candidate for studying the quantum mechanics of macroscopic oscillators [34].

## II. EXPERIMENTAL DETAIL

The experimental schematic is shown in the Fig. 1(b). The cavity and magnet are placed on the base plate of the dilution refrigerator where the temperature can be reduced to ~50 mK, which is below the zero-field critical temperature of aluminum ($T_C$~1.2 K), while the vacuum pressure is held to ~$10^{-7}$ mbar. A pin antenna is coupled to the quarter-wave SRF mode at the stub height inside the cavity, where the electric field is strongly localized. Measurements are performed using a vector network analyzer whose probe signal passes through a circulator, is reflected by the cavity, and the reflection is separated by the circulator. For these measurements the cavity is over-coupled which reduces the loaded Q, but allows us to measure the resonance frequency from room temperature down to 50 mK. The cavity is held in place with N-type cryo grease to ensure thermal conductivity with the cryostat. We conducted a sequence of three experiments. First, the cavity (including a plastic sleeve around the stub whose purpose is to keep the magnet from falling off the top of the stub) is cooled without any magnet present. Here we see the dependence of the cavity resonance frequency as a function of temperature. Second, the magnet is placed within the cavity on the top of the stub where the plastic sleeve keeps the magnet from falling off of the top of the stub. Third, the same magnet is placed at the bottom of the same cavity. In all of these experiments all other experimental conditions remain the same.

The dimension and shape of the cavity are shown in the Fig. 1 (a). This cavity is made up by 6061 aluminum. It is a coaxial cylindrical cavity with one end open. The outer cylinder has radius and height, respectively, 7 mm and 55 mm. The inner cylinder has radius of 2 mm and height of 5 mm giving a quarter-wave resonance of 10 GHz. In our design, the thickness of the metal on the bottom part of the cavity is 6.5 mm. The cylindrical neodymium magnet (shown in the Fig. 1 (c)) is 1 mm high and has a radius of 0.375 mm. The N50 magnet has a remanence of 1.44 T (provided by manufacturer) and a mass of 4 mg. Assuming homogeneous magnetization, the magnetic dipole moment is $5\times10^{-4}$ A.m$^2$.

When the cavity is cooled to below its superconducting transition, we expect the magnet to levitate above the surface due to the Meissner effect in Type I superconductors. Fig. 1 (d) illustrates the expected changes to the resonance frequency of the cavity during levitation experiments. The frequency, $f$, of a coaxial stub cavity is determined by the height of the stub, $l$, where $f \alpha \frac{4}{l}$ [2]. Any perturbation within the coaxial region of the cavity changes the shape of the cavity mode and hence its frequency [35]. When a magnet is placed on the surface of the stub, it increases effective height of the stub and hence decreases frequency of the cavity. The amount of this downshift corresponds to the interaction of the magnet with the electric field of the cavity mode, which is concentrated toward the edges of the stub. Conversely, when the magnet is placed on the bottom of the cavity it raises floor of the cavity. This reduces the effective length of the stub which causes the resonance frequency to increase. As the magnet levitates above either surface of the cavity, the frequency of the cavity shifts towards the frequency of the bare cavity (cavity without any magnet).

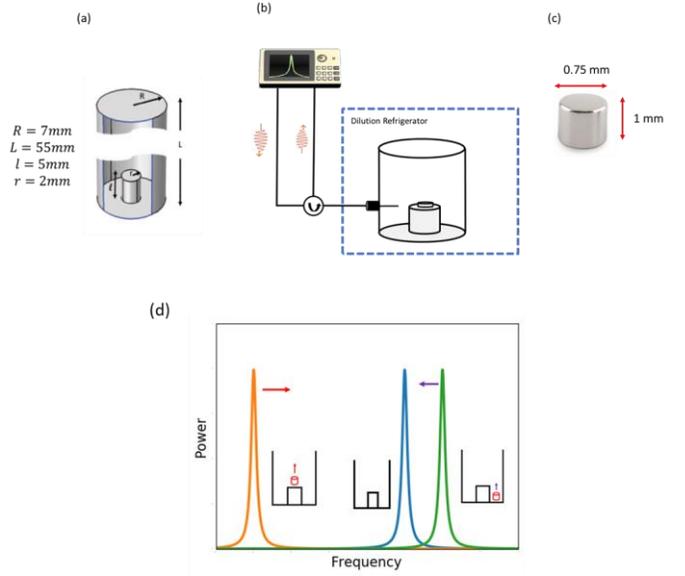

Fig. 1. (a) This is a schematic of the the cavity used in our work. It has a coaxial part, which contains a stub, and a cylindrical part has one open end and is closed at the other. (b) Schematic of the experimental set up. A magnet is placed on the top part of the stub of the coaxial microwave cavity. The cavity is probed by the signal sent from the network analyzer. (c) The shape and size of the N50 magnet used in our work. It has a maximum field strength of 1. 44 T and surface field of 0.67 T. (d) Expected frequency shift with the position of the magnet. When there is no magnet in the cavity (a bare cavity), the resonance frequency is fixed and given by the blue curve. If the magnet rests on the bottom of the cavity, the cavity resonance frequency is higher than that of the bare cavity as shown by the green curve. As the magnet levitates above the bottom of the cavity, we expect the resonance frequency to shift lower as a function of levitation height (as indicated by the left-pointing purple arrow at the tip of the green curve). If the magnet rests on the top of the stub, its resonance frequency is significantly lower than that of the bare cavity as shown by the orange curve. As the magnet levitates above the stub, we expect the resonance frequency to shift higher as a function of levitation height (as indicated by the right-pointing red arrow at the tip of the orange curve). The sensitivity of the resonance frequency as a function of position depends on the radial position of the magnet. In both cases, as the magnet levitates the resonance frequency shifts towards that of the bare cavity.

## III. ROOM TEMPERATURE MEASUREMENTS AND SUMILATION

In order to gain an understanding of the expected frequency shifts when levitation occurs, we measured the resonance frequency as a function of the position of a magnet in the cavity at room temperature. The magnet was held in a dielectric capillary and its position was controlled using micrometer stages. Measurements were also made with just the capillary and no magnet to eliminate the effect of the capillary on the measurements. Fig. 2 shows the resonance frequency shift as a function of the gap between the magnet and the tip of the stub. The solid curves are simulations and the symbols are experimental data. In particular, we expect the magnitude of the frequency downshift to be between -100 MHz and -200 MHz when the magnet is in contact with the surface of the stub. The resonance frequency comes within 5 MHz of that of the bare cavity once the magnet height is greater than 3 mm. We observe stronger coupling when the magnet is located near the edge of the stub because the electric field is concentrated around the edge. Finite element calculations (COMSOL Multiphysics) reveal the same trends. In these calculations, the size of the magnet and the cavity are the same as the experimental work. These room temperature measurements and simulations allow us to predict the behavior of the system during levitation experiments.

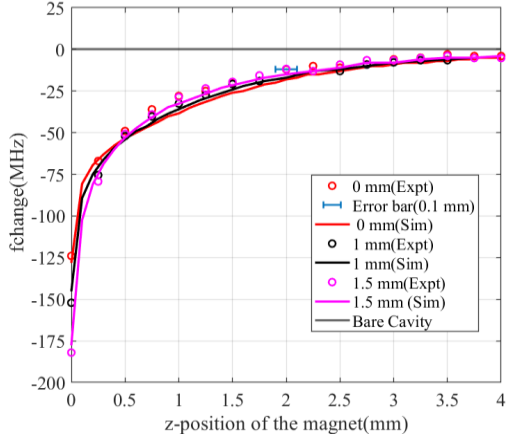

Fig. 2. Measurements are taken at room temperature by putting a magnet inside a capillary tube and by sealing its end by a tape, and factoring out the effect of the capillary. The capillary tube with the magnet is positioned at different coordinates inside the cavity by the translational stage of the micrometer. The error bar is same for all experimental data.

## IV. LOW TEMPERATURE MEASUREMENTS

The frequency of the bare cavity without the magnet is 10.041 GHz at 5K, shifting upward by 5 kHz as the temperature decreases from 5 K to 55 mK. With the bare cavity frequency as a reference, when the magnet is placed on the stub, the frequency shifts down by 130 MHz to 9.910 GHz. The amount of frequency shift varies with the exact position of the magnet on the tip of the stub with the maximum shift occurring when the magnet is placed near the edge of the stub. For the case of a magnet on the stub, we observed 120 MHz of frequency change as the temperature drops from 5 K to 65 mK, as shown by the solid lines in Fig. 3(b). The sudden frequency shift at 650 mK is consistent with levitation due to expulsion of the magnetic field from the material below the magnet. As the temperature drops below 100 mK the frequency shift approaches -12 MHz. Our measurements are consistent over several heating and cooling cycles. We also report measurements for the case where the magnet is placed in the bottom of the cavity. The variations in the cavity resonance frequency are smaller, but still measurable.

At 5 K the difference between frequency of bare cavity and cavity with magnet on the bottom is +11 MHz. As temperature is reduced to 57 mK the frequency difference drops to +5 MHz.

The neodymium permanent magnet has a maximum magnetic field (remanence) of 1.44 T and a surface field of 0.67 T.

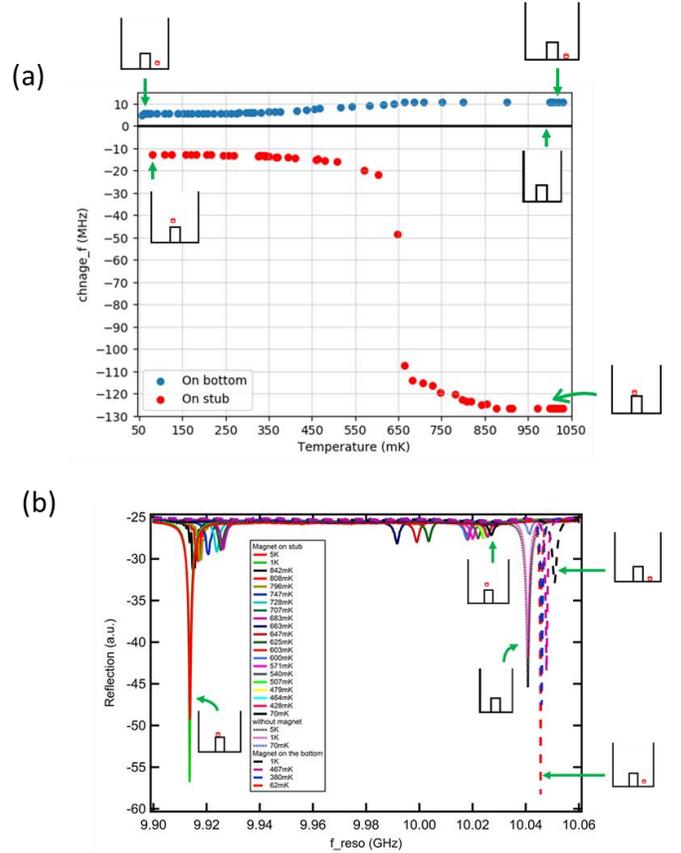

Fig. 3. (a): Cavity resonance frequency shift as a function of temperature. (b): A composite of reflection spectrum measurements for the three cavity configurations and for a range of temperatures. When a magnet is located on the top of the stub and cooled to 70mK, we observe an downward shift in frequency (solid lines) which is reduced from -130 MHz at 1 K down to -12 MHz at 70 mK. When the magnet is located on the bottom of the cavity, the frequency is up shifted and drops from 11 MHz at 1 K to 5 MHz at 50 mK (dashed lines).

When the magnet rests directly on the surface of the aluminum, the magnetic field strength at the interface is larger than the critical field of the aluminum 100 gauss at 1.2 K. This creates a normal conducting region having a depth of 1-2 mm directly below the magnet. As the temperature drops, the critical field increases and the normal region becomes thinner. Below 650 mK there is sufficient lift due to the Meissner effect to offset the gravitational force and levitate the magnet. Once the magnet begins to lift at 650 mK, the normal region shrinks which increases the Meissner force making magnet jump up to its equilibrium position of 2.5 mm. At this point the magnetic flux is completely expelled from the volume of the superconductor. For the case of the magnet levitated 2.5 mm above the stub, the sensitivity of the resonance frequency to height fluctuations is 10 MHz/mm. For levitation above the bottom of the cavity we expect the sensitivity to be less than above the stub by a factor of 50.

## V. CONCLUSION

We demonstrate levitation of a millimeter-sized neodymium permanent magnet to a height of 2.5 mm within the interior of a SRF cavity. The levitation temperature is 650 mK and we find that the levitation above the stub exhibits a larger sensitivity compared with levitation from the bottom of the cavity. Our future goals include characterizing the mechanical vibrations of the levitated magnet and integrating this magnet-cavity system with other cavity optomechanical schemes.

## ACKNOWLEDGMENT

N. K. Raut would like to thank Dr. Luis Martinez, Dr. Alessandro Castelli. The author would also like to thank Sharping group in UC Merced.